\documentclass{article}
\usepackage{amsmath,amssymb,graphicx,psfrag,theorem}

\newcommand{\smfrac}[2]{\ensuremath{\tfrac{#1}{#2}}}
\newcommand{\leg}[2]{\ensuremath{\genfrac{(}{)}{}{}{#1}{#2}}}
\newcommand{\smleg}[2]{\ensuremath{\genfrac{(}{)}{}{1}{#1}{#2}}}

\newcommand{\subphi}{{\scriptscriptstyle {\ket{\phi}}}}
\newcommand{\subtphi}{{\scriptscriptstyle {\ket{\tilde{\phi}}}}}
\newcommand{\distD}{{\mathcal D}}
\newcommand{\rf}{{\mbox{\tiny RF}}}
\newcommand{\cf}{{\mbox{\tiny CF}}}

\newcommand{\Z}{\ensuremath{\mathbb{Z}}}
\newcommand{\C}{\ensuremath{\mathbb{C}}}
\newcommand{\F}{{\ensuremath{\mathbb F}}}
\newcommand{\Zr}[1]{\ensuremath{\mathbb{Z}/{#1}\mathbb{Z}}}
\newcommand{\om}[2]{\omega_{#1}^{#2}}
\newcommand{\ket}[1]{\ensuremath{|#1\rangle}}

\newcommand{\Tr}{{\ensuremath{\mathrm{Tr}}}}
\newcommand{\Fp}{\ensuremath{\mathbb{F}_p}}
\newcommand{\Fq}{\ensuremath{\mathbb{F}_q}}
\newcommand{\FT}{\ensuremath{\mathcal{F}}}

\newtheorem{lemma}{Lemma}
\newenvironment{proof}{\noindent{\bf Proof: }} {\hfill\rule{2mm}{2mm} \newline}
\newtheorem{definition}{Definition}
{\theorembodyfont{\rmfamily} \newtheorem{algorithm}{Algorithm}}
\newtheorem{theorem}{Theorem}
\newtheorem{corollary}{Corollary}

\begin{document}

\title{\textbf{\Large Quantum Algorithms for some Hidden Shift Problems}}

\author{Wim van Dam\thanks{HP Labs, Palo Alto; Mathematical Sciences
    Research Institute; and Computer Science Division, University of
    California, Berkeley.  Supported by an HP - MSRI postdoc
    fellowship.
  } \\
  HP, MSRI, U.C. Berkeley\\
  vandam@cs.berkeley.edu \and Sean Hallgren\thanks{Supported in part
    by an NSF Mathematical Sciences Postdoctoral Fellowship and in
    part by the NSF through the Institute for Quantum Information at
    the California Institute of Technology.  Most of this work done
    while the author was at the Mathematical Sciences Research
    Institute and the University of California, Berkeley, with partial
    support from DARPA QUIST Grant F30602-01-2-0524.
  } \\
  Caltech\\
  hallgren@cs.caltech.edu \and Lawrence Ip\thanks{Computer Science
    Division, University of California, Berkeley.  Supported by NSF
    Grant CCR-0049092, DARPA Grant F30602-00-2-0601 and DARPA QUIST
    Grant F30602-01-2-2054.  Part of this work was done while the
    author was a visitor at the Institute for Quantum Information at
    the California Institute of Technology.
  }\\
  U.C. Berkeley\\
  lip@eecs.berkeley.edu
}
\date{}

\maketitle

\begin{abstract}
\small
 Almost all of the most successful quantum algorithms discovered
 to date
 exploit the ability of the Fourier transform to recover
 subgroup structure of functions, especially periodicity.
 The fact that Fourier transforms can also be used to capture
 shift structure has received far less attention
 in the context of quantum computation.

\small
 In this paper, we present three examples of ``unknown shift'' problems
 that can be solved efficiently on a quantum computer using
 the quantum Fourier transform.
 We also define the \emph{hidden coset problem,} which generalizes
 the hidden shift problem and the hidden subgroup problem.
 This framework provides a unified way of viewing the ability of
 the Fourier transform to capture subgroup and shift structure.
\end{abstract}

\section{Introduction}

The first problem to demonstrate a superpolynomial separation
between random and quantum polynomial time was the Recursive
Fourier Sampling problem~\cite{SICOMP::BernsteinV1997}.
Exponential separations were subsequently discovered by
Simon~\cite{SICOMP::Simon1997:1474}, who gave an oracle problem,
and by Shor~\cite{Shor}, who found polynomial time quantum
algorithms for factoring and discrete log. We now understand that
the natural generalization of Simon's problem and the factoring
and discrete log problems is the hidden subgroup problem (HSP),
and that when the underlying group is Abelian and finitely
generated, we can solve the HSP efficiently on a quantum computer.
While recent results have continued to study important
generalizations of the HSP (for example,~\cite{EttingerHK99,
STOC::HallgrenRT2000, GSVV, Watrous, IMSantha,
  Hallgren2002}), only the Recursive Fourier Sampling problem remains
outside the HSP framework.

In this paper, we give quantum algorithms for several hidden shift
problems. In a hidden shift problem we are given two functions
$f$, $g$ such that there is a shift $s$ for which $f(x)=g(x+s)$
for all $x$. The problem is then to find $s$. We show how to solve
this problem for several classes of functions, but perhaps the
most interesting example is the shifted Legendre symbol problem,
where $g$ is the
  Legendre symbol\footnote{The Legendre symbol
  $\smleg{x}{p}$ is defined to be 0 if $p$ divides $x$, 1 if $x$ is a
  quadratic residue mod $p$ and $-1$ if $x$ is not a quadratic residue
  mod $p$.} with respect to a prime
size finite field, and the problem is then: ``Given the function
$f(x)=\smleg{x+s}{p}$ as an oracle, find $s$''.

The oracle problem our algorithms solve can be viewed as the
problem of predicting a pseudo-random function $f$. Such tasks
play an important role in cryptography and have been studied
extensively under various assumptions about how one is allowed to
query the function (nonadaptive versus adaptive, deterministic
versus randomized, et cetera)~\cite{BlumMicali,crypto}. In this
paper we consider the case where the function is queried in a
quantum mechanical superposition of different values $x$. We show
that if $f(x)$ is an $s$-shifted multiplicative character
$\chi(x+s)$, then a polynomial-time quantum algorithm making such
queries can determine the hidden shift $s$, breaking the
pseudo-randomness of $f$.  We conjecture that classically the
shifted Legendre symbol is a pseudo-random function, that is, it
is impossible to efficiently predict the value of the function
after a polynomial number of queries if one is only allowed a
classical algorithm with oracle access to $f$. Partial evidence
for this conjecture has been given by Damg{\aa}rd~\cite{Damgard}
who proposed the related task: ``Given a part of the Legendre
sequence $\smleg{s}{p},\smleg{s+1}{p},\dots,\smleg{s+\ell}{p}$,
where $\ell$ is $O(\log p)$, predict the next value
$\smleg{s+\ell+1}{p}$'', as a hard problem with applications in
cryptography.

Using the quantum algorithms presented in this paper, we can break
certain algebraically homomorphic cryptosystems by a reduction to
the shifted Legendre symbol problem. The best known classical
algorithm~\cite{BonehLiptonBBF} for breaking these cryptosystems
is subexponential and is based on a smoothness assumption.  These
cryptosystems can also be broken by Shor's algorithm for period
finding, but the two attacks on the cryptosystems appear to use
completely different ideas.

While current quantum algorithms solve problems based on an underlying
group and the Fourier transform over that group, we initiate the study
of problems where there is an underlying ring or field.  The Fourier
transform over the additive group of the ring is defined using the
characters of the additive group, the additive characters of the ring.
Similarly, the multiplicative group of units induces multiplicative
characters of the ring. The interplay between additive and
multiplicative characters is well
understood~\cite{LidlNiederreiter,TolimieriAnLu}, and we show that
this connection can be exploited in quantum algorithms. In particular,
we put a multiplicative character into the phase of the registers and
compute the Fourier transform over the additive group. The resulting
phases are the inner products between the multiplicative character and
each of the additive characters, a Gauss sum. We hope the new tools
presented here will lead to other quantum algorithms.

We give algorithms for three types of hidden shift problems:

In the first problem, $g$ is a multiplicative character of a
finite field. Given $f$, a shifted version of $g$, the shift is
uniquely determined from $f$ and $g$. An example of a
multiplicative character of ${\Zr{p}}$ is the Legendre symbol. Our
algorithm uses the Fourier transform over the additive group of a
finite field.

In the second problem, $g$ is a multiplicative character of the
ring $\Zr{n}$. This problem has the feature that the shift is not
uniquely determined by $f$ and $g$ and our algorithm identifies
all possible shifts. An example of a multiplicative character of
$\Zr{n}$ is the Jacobi symbol\footnote{The Jacobi symbol
$\smleg{a}{b}$ is defined so that it satisfies the relation
$\smleg{a}{bc} = \smleg{a}{b} \smleg{a}{c}$ and reduces to the
Legendre symbol when the lower parameter is prime.}.

In the third problem we have the same setup as in the second
problem with the additional twist that $n$ is unknown.

We also define the \emph{hidden coset problem}, which is a
generalization of the hidden shift problem and the hidden subgroup
problem.  This definition provides a unified way of viewing the
quantum Fourier transform's ability to capture subgroup and shift
structure.

Some of our hidden shift problems can be reduced to the HSP,
although efficient algorithms for these HSP instances are not
known. Assuming Conjecture~2.1 from~\cite{BonehLiptonBBF}, the
shifted Legendre symbol problem over \Zr{p} can be reduced to an
instance of the HSP over the dihedral group $D_p=\Zr{p} \rtimes
\Zr{2}$ in the following way.  Let $f(x,0) = (\leg{x}{p},
\leg{x+1}{p}, \dots, \leg{x+\ell}{p})$ and $f(x,1) =
(\leg{x+s}{p}, \leg{x+s+1}{p}, \dots, \leg{x+s+\ell}{p})$, where
$s$ is unknown and $\ell > 2 \log^2 p$. Then the hidden subgroup
is $H=\{(0,0),(s,1)\}$.  This conjecture is necessary to ensure
that $f$ will be distinct on distinct cosets of $H$.  For the
general shifted multiplicative character problem, the analogous
reduction to the HSP may fail because $f$ may not be distinct on
distinct cosets. However, we can efficiently generate random coset
states, that is, superpositions of the form $\ket{x,0} +
\ket{x+s,1}$, although it is unknown how to use these to
efficiently find~$s$~\cite{EttingerHoyer}. The issue of
nondistinctness on cosets in the HSP has been studied for some
groups~\cite{BonehLiptonHiddenLinearFunctions,HalesHallgren,Hales02,FMSS}.

The existence of a time efficient quantum algorithm for the
shifted Legendre symbol problem was posed as an open question
in~\cite{weighingmatrix}.  The Fourier transform over the additive
group of a finite field was independently proposed for the
solution of a different problem in~\cite{deBeaudrapCleveWatrous}.
The current paper subsumes~\cite{vandamhallgren} and~\cite{Ip}.
Building on the ideas in this paper, a quantum algorithm for
estimating Gauss sums is described in~\cite{DamSeroussi}.

This paper is organized as follows. Section~\ref{sect:background}
contains some definitions and facts. In Section~\ref{sect:idea},
we give some intuition for the ideas behind the algorithms. In
Section~\ref{sect:finitefields}, we present an algorithm for the
shifted multiplicative problem over finite fields, of which the
shifted Legendre symbol problem is a special case, and show how we
can use this algorithm to break certain algebraically homomorphic
cryptosystems. In Section~\ref{sect:rings}, we extend our
algorithm to the shifted multiplicative problem over rings
$\Zr{n}$. This has the feature that unlike in the case of the
finite field, the possible shifts may not be unique. We then show
that this algorithm can be extended to the situation where $n$ is
unknown. In Section~\ref{sect:HCP}, we show that all these
problems lie within the general framework of the hidden coset
problem. We give an efficient algorithm for the hidden coset
problem provided $g$ satisfies certain conditions. We also show
how our algorithm can be interpreted as solving a deconvolution
problem using Fourier transforms.

\section{Background}
\label{sect:background}

\subsection{Notation and Conventions}
We use the following notation:
 $\om{n}{}$ is the $n$th root of unity $\exp(2 \pi i /n)$,
and $\hat{f}$ denotes the Fourier transform of the function $f$.
 An algorithm computing in $\F_q$, $\Zr{n}$ or $G$ runs in
 polynomial time if it runs in time polynomial in
 $\log q$, $\log n$ or $\log|G|$.

 In a ring $\Zr{n}$ or a field $\F_q$, additive characters $\psi$
($\Zr{n}\to \C^*$ or $\F_q \to \C^*$) are characters of the
additive group, that is, $\psi(x+y)=\psi(x)\psi(y)$, and
multiplicative characters $\chi$ ($(\Zr{n})^*\to \C^*$ or
$\F_q^*\to \C^*$) are characters of the multiplicative group of
units, that is, $\chi(xy)=\chi(x)\chi(y)$ for all $x$ and $y$. We
extend the definition of a
 multiplicative character to the entire ring or field by assigning
 the value zero to elements outside the unit group.
All nonzero $\chi(x)$ values have unit norm and so
$\chi(x^{-1})=\overline{\chi(x)}$.

We ignore the normalization term in front of a superposition
unless we need to explicitly calculate the probability of
measuring a particular value.

\subsection{Computing Superpositions}

We will need to compute the superposition $\sum_{x} f(x)\ket{x}$
where $f(x)$ is in the \emph{amplitude}.
\begin{lemma}[Computing Superpositions]
 \label{lemma:superposition}
 Let $f:G \to \C$ be a complex-valued function defined on
 the set $G$ such that $f(x)$ has unit magnitude whenever $f(x)$ is
 nonzero. Then there is an efficient algorithm for creating
 the superposition
 $\sum_{x}f(x)\ket{x}$ with success probability equal to the fraction of
 $x$ such that $f(x)$ is nonzero and that uses only two queries to the
function $f$.
\end{lemma}

\begin{proof}
Start with the superposition over all $x$, $\sum_{x} \ket{x}$.
Compute $f(x)$ into the second register and measure to see whether
$f(x)$ is nonzero. This succeeds with probability equal to the
fraction of $x$ such that $f(x)$ is nonzero. Then we are left with
a superposition over all $x$ such that $f(x)$ is nonzero. Compute
the phase of $f(x)$ into the phase of \ket{x}. This phase
computation can be approximated arbitrarily closely by
approximating the phase of $f(x)$ to the nearest $2^n$th root of
unity for sufficiently large~$n$.  Use a second query to $f$ to
reversibly uncompute the $f(x)$ from the second register.
\end{proof}

\subsection{Approximate Fourier Sampling}

\label{sect:background:FourierSampling}

It is not known how to efficiently compute the quantum Fourier
transform over $\Zr{n}$ exactly. However, efficient approximations
are known~\cite{Kitaev1995,Kitaev1997,CleveFourier,HalesHallgren}.
We can even compute an efficient approximation to the distribution
induced when $n$ is unknown as long as we have an upper bound on
$n$~\cite{HalesHallgren}. We will need to approximately Fourier
sample to solve the unknown $n$ case of the shifted character
problem in Section~\ref{sect:unknown}.

To Fourier sample a state $\ket{\phi}$, we form the state
$\ket{\tilde{\phi}}$ that is the result of repeating $\ket{\phi}$
many times. We then Fourier sample from $\ket{\tilde{\phi}}$ and
use continued fractions to reduce the expanded range of values.
This expansion into $\ket{\tilde{\phi}}$ allows us to perform the
Fourier sampling step over a length from which we \emph{can}
exactly Fourier sample.

More formally, let $\ket{\phi} = \sum_{x=0}^{n-1}{\phi_x \ket{x}}$
be an arbitrary superposition, and $\hat{\distD}_\subphi$ be the
distribution induced by Fourier sampling $\ket{\phi}$ over $\Z_n$.
Let the superposition $\ket{\tilde{\phi}} =
\sum_{x=0}^{m-1}{\phi_{x \bmod n}
  \ket{x}}$ be $\ket{\phi}$ repeated until some arbitrary integer $m$,
not necessarily a multiple of $n$.  Let 
$\hat{\distD}_\subtphi$ 
be the distribution induced by Fourier sampling $\ket{\tilde{\phi}}$
over $\Z_q$ rather than $\Z_m$ (where $q>m$
 and $\phi_x=0$ if $x\geq m$).  Notice that
$\hat{\distD}_\subphi$ is a distribution on $\Z_n$ and
$\hat{\distD}_\subtphi$ is a distribution on $\Z_q$.

We can now define the two distributions we will compare.  Let
$\hat{\distD}_\subphi^\rf$ be the distribution induced on the
reduced fractions of $\hat{\distD}_\subphi$, that is, if $x$ is a
sample from $\hat{\distD}_\subphi$, we return the fraction $x/n$
in lowest terms.  In particular, define
$\hat{\distD}_\subphi^\rf(j,k)=\hat{\distD}_\subphi(jm)$ if
$mk=n$. Let $\hat{\distD}_\subtphi^\cf$ be the distribution
induced on fractions from sampling $\hat{\distD}_\subtphi$ to
obtain $x$, and then using continued fractions to compute the
closest approximation to $x/q$ with denominator at most~$n$. If
$m= \Omega(\frac{n^2}{\epsilon^2})$ and $q
=\Omega(\frac{m}{\epsilon})$, then $|\hat{\distD}_\subphi^\rf -
\hat{\distD}_\subtphi^\cf |_1< \epsilon$.

\subsection{Finite Fields}

The elements of a finite field $\F_q$ (where $q=p^r$ for some
prime $p$) can be represented as polynomials in $\F_p[X]$ modulo
a degree $r$ irreducible polynomial in $\F_p[X]$. In this
representation, addition, subtraction, multiplication and
division can all be performed in $O((\log q)^2)$
time~\cite{BachShallit}.

We will need to compute the Fourier transform over the additive
group of a finite field, which is isomorphic to $(\Zr{p})^r$.
 The additive characters are of the form
$\psi_y(x)=\om{p}{\Tr(xy)}$, where $\Tr:\Fq\to\Fp$ is the trace of
the finite field $\Tr(x) = \sum_{j=0}^{r-1}{x^{p^j}}$, and
$y\in\Fq$~\cite{LidlNiederreiter}. We can efficiently compute the
Fourier transform over the additive group of a finite field.
\begin{lemma}[Fourier Transform over $\Fq$] \label{thm:TFT}
 \newcounter{appendixlemma}
 \setcounter{appendixlemma}{\value{lemma}}
  The Fourier transform $\ket{x} \mapsto \frac{1}{\sqrt{q}}
  \sum_{y\in \Fq} \om{p}{\Tr(xy)}\ket{y}$
  can be approximated to within error $\epsilon$ in time
  polynomial in $\log q$ and $\log 1/\epsilon$.
\end{lemma}
\begin{proof}
See~\cite{vandamhallgren}.
(Independently, the efficiency of this transform was also shown
in~\cite{deBeaudrapCleveWatrous}.)
\end{proof}
For clarity of exposition we assume throughout the rest of the
paper that this Fourier transform can be performed exactly, as we
can make the errors due to the approximation exponentially small
with only polynomial overhead.

\subsection{Multiplicative Characters and their Fourier
Transforms}

The multiplicative group $\Fq^*$ of a finite field $\Fq$ is
cyclic. Let $g$ be a generator of $\Fq^*$. Then the multiplicative
characters of $\Fq$ are of the form $\chi(g^\ell) =
\om{q-1}{k\ell}$ for all $\ell \in \{0, \dots, q-2\}$ where the
$q-1$ different multiplicative characters are indexed by $k \in
\{0, \dots, q-2\}$. The trivial character is the character with
$k=0$. We can extend the definition of $\chi$ to $\Fq$ by defining
$\chi(0) = 0$. On a quantum computer we can efficiently compute
$\chi(x)$ because the value is determined by the discrete
logarithm $\log_g(x)$, which can be computed efficiently using
Shor's algorithm~\cite{Shor}. The Fourier transform of a
multiplicative character $\chi$ of the finite field $\Fq$ is given
by $\hat{\chi}(y)=\overline{\chi(y)}
\hat{\chi}(1)$~\cite{LidlNiederreiter,TolimieriAnLu}.

Let $n=p_1^{m_1}\dots p_k^{m_k}$ be the prime factorization of
$n$. Then by the Chinese Remainder Theorem, $(\Zr{n})^* \cong
(\Zr{p_1^{m_1}})^* \times \cdots \times (\Zr{p_k^{m_k}})^*$. Every
multiplicative character $\chi$ of $\Zr{n}$ can be written as the
product $\chi(x) = \chi_1(x_1) \dots \chi_k(x_k)$, where $\chi_i$
is a multiplicative character of $\Zr{p_i^{m_i}}$ and $x_i \equiv
x \bmod p_i^{m_i}$.  We say $\chi$ is \emph{completely nontrivial}
if each of the $\chi_i$ is nontrivial. We extend the definition of
$\chi$ to all of $\Zr{n}$ by defining $\chi(y)=0$ if
$\gcd(y,n)\neq 1$. The character $\chi$ is aperiodic on
$\{0,\dots,n-1\}$ if and only if all its $\chi_i$ factors are
aperiodic over their respective domains $\{0,\dots,p_i^{m_i}-1\}$.
We call $\chi$ a \emph{primitive character} if it is completely
nontrivial and aperiodic. Hence, $\chi$ is primitive if and only
if all its $\chi_i$ terms are primitive.

It is well known that the Fourier transform of a primitive $\chi$
is $\hat{\chi}(y)=\overline{\chi(y)}\hat{\chi}(1)$.  If $\chi$ is
completely nontrivial but periodic with period $\ell$, then its
Fourier transform obeys $\hat{\chi}(yn/\ell) =
\overline{\chi'(y)}\hat{\chi}'(1)$, where $\chi'$ is the primitive
character obtained by restricting $\chi$ to $\{0,\dots,\ell-1\}$.
See the book by Tolimieri et al.\ for
details~\cite{TolimieriAnLu}.

\section{Intuition Behind the Algorithms for the Hidden Shift Problem}
\label{sect:idea}

We give some intuition for the ideas behind our algorithms for the
hidden shift problem.  We use the shifted Legendre symbol problem
as our running example, but the approach works more generally. In
the shifted Legendre symbol problem we are given a function $f_s:
\Z_p \to \{0,\pm 1\}$ such that $f(x)=\smleg{x+s}{p}$, and are
asked to find $s$. The Legendre symbol $\smleg{\cdot}{p}:\F_p\to
\{0,\pm 1\}$ is the quadratic multiplicative character of $\F_p$
defined: $\smleg{x}{p}$ is $1$ if $x$ is a square modulo $p$, $-1$
if it is not a square, and $0$ if $x \equiv 0$.

The algorithm starts by putting the function value in the phase to
get $\ket{f_s} = \sum_{x}{f_s(x)\ket{x}} = \sum_{x} \leg{x+s}{p}
\ket{x}$. Assume the functions $f_z$ are mutually (near)
orthogonal for different $z$, so that the inner product $\langle
f_z | f_s \rangle$ approximates the delta function value
$\delta_{s}(z)$.  Using this assumption, define the (near) unitary
matrix $C$, where the $z$th row is $\ket{f_z}$. Our quantum state
$\ket{f_s}$ is one of the rows, hence $C \ket{f_s} = \ket{s}$.
The problem then reduces to: how do we efficiently implement $C$?
By definition, $C$ is a circulant matrix ($c_{x,y} =
c_{x+1,y+1}$).  Since the Fourier transform matrix diagonalizes a
circulant matrix, we can write $C = \FT (\FT^{-1} C \FT ) \FT^{-1}
= \FT D \FT^{-1}$, where $D$ is diagonal.  Thus we can implement
$C$ if we can implement $D$.  The vector on the diagonal of $D$ is
the vector $\FT^{-1}\ket{f_0}=\FT^{-1} \sum_x \leg{x}{p} \ket{x}$,
the inverse Fourier transform of the Legendre symbol. The Legendre
symbol is an eigenvector of the Fourier transform, so the diagonal
matrix contains the values of the Legendre symbol times a global
constant that can be ignored. Because the Legendre symbol can be
computed efficiently classically, it can be computed into the
phase, so $C$ can be implemented efficiently.

In summary, to implement $C$ for the hidden shift problem for the
Legendre symbol, compute the Fourier transform, compute
$\smleg{x}{p}$ into the phase at $\ket{x}$, and then compute the
Fourier transform again (it is not important whether we use $\FT$
or $\FT^{-1}$).

\begin{figure*}
\begin{center}
 \psfrag{|0>}[cl]{\small \ket{0}}
 \psfrag{FT}[cc]{$\FT$}
 \psfrag{ftophase}[cc]{\small $\ket{x} \mapsto f(x)\ket{x}$}
 \psfrag{gtophase}[cc]{\small $\ket{x} \mapsto \hat{g}^{-1}(x)\ket{x}$}
 \psfrag{measure}[cl]{measure}
 \scalebox{1}{\includegraphics{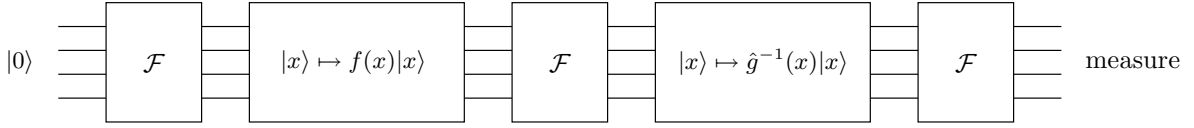}}
 \caption{Circuit for hidden shift problem.
 Notice how we compute $f$ and $\hat{g}^{-1}$
into the \emph{phase}.}
 \label{fig:shift}
\end{center}
\end{figure*}

\begin{figure*}
\begin{center}
 \psfrag{f}[cc]{\small $\ket{x} \mapsto \ket{x}\ket{f(x)}$}
 \psfrag{|0>}[cl]{\small \ket{0}}
 \psfrag{FT}[cc]{$\FT$}
 \psfrag{measure}[cl]{measure}
 \scalebox{1}{\includegraphics{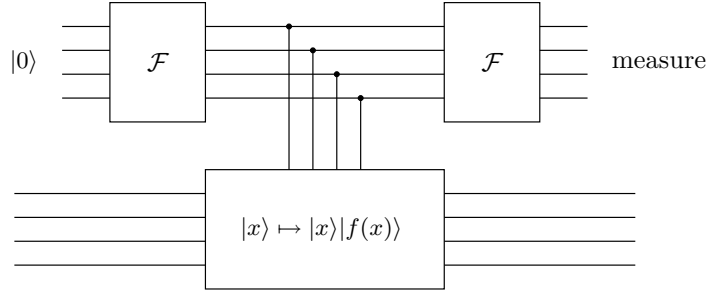}}
 \caption{Circuit for hidden subgroup problem.
 Here $f$ is computed into a \emph{register}.}
 \label{fig:hsp}
\end{center}
\end{figure*}

Figure~\ref{fig:shift} shows a circuit diagram outlining the
algorithm for the hidden shift problem in general. Contrast this
with the circuit for the hidden subgroup problem shown in
Figure~\ref{fig:hsp}.

\section{Shifted Multiplicative Characters of Finite Fields}
\label{sect:finitefields}

In this section we show how to solve the hidden shift problem for
any nontrivial multiplicative character of a finite field. The
Fourier transform we use is the Fourier transform over the
additive group of the finite field.

\begin{definition}\textsc{(Shifted Multiplicative Character Problem
over Finite Fields)} Given a nontrivial multiplicative character
$\chi$ of a finite field~$\F_q$  (where $q=p^r$ for some prime
$p$), and a function $f$ for which there is an $s$ such that
$f(x)=\chi(x+s)$ for all $x$. Find $s$.
\end{definition}

\begin{algorithm}\textsc{(Shifted Multiplicative Character Problem
 over Finite Fields)}
  \label{alg:finitefield}
\ 
  \begin{enumerate}
    \item Create $\sum_{x\in \F_q} \chi(x+s) \ket{x}$.
      \label{alg:finitefield:superposition}
    \item Compute the Fourier transform to obtain
      $\sum_{y\in \F_q}\om{p}{\Tr(-sy)}\hat{\chi}(y) \ket{y}$.
      \label{alg:finitefield:1stFT}
    \item For all $y\neq 0$, compute $\chi(y)$
      into the phase to obtain
      $\hat{\chi}(1)\sum_{y\in \F_q^*} \om{p}{\Tr(-sy)} \ket{y}$.
      \label{alg:finitefield:conjugate}
    \item Compute the inverse Fourier transform and measure the outcome
       $-s$.
      \label{alg:finitefield:2ndFT}
  \end{enumerate}
\end{algorithm}

\begin{theorem}
  \label{thm:ff}
  For any finite field and any nontrivial multiplicative character,
  Algorithm~\ref{alg:finitefield} solves the shifted multiplicative
  character problem over finite fields with
  probability $(1-1/q)^2$.
\end{theorem}
\begin{proof}
\begin{enumerate}
 \item
  Since $\chi(x)=0$ only
  at $x=0$, by Lemma~\ref{lemma:superposition} we can create the
  superposition with probability $1-1/q$.
 \item
  By Lemma~\ref{thm:TFT} we can compute the Fourier transform
  efficiently.
  The Fourier transform moves the shift~$s$ into the phase as
  described.
 \item Because
  $\hat{\chi}(y) =\overline{\chi(y)}\hat{\chi}(1)$ for every nonzero $y$,
  the phase change $\ket{y}\mapsto \chi(y)\ket{y}$
  establishes the required transformation.
 \item
  The amplitude of \ket{-s} is \newline
  $\frac{1}{\sqrt{q}}\frac{1}{\sqrt{q-1}}
    \sum_{y\in\F_q^*}\om{p}{\Tr(-sy)}\om{p}{\Tr(sy)}$
     $=\frac{1}{\sqrt{q}}\frac{1}{\sqrt{q-1}}\sum_{y\in\F_q^*} 1$
     $=\sqrt{\frac{q-1}{q}}$,
  so the probability of measuring $-s$ is $1-1/q$.
\end{enumerate}
\end{proof}

\subsection{Example: The Legendre Symbol and Homomorphic
Encryption}

The Legendre symbol $\smleg{\cdot}{p}:\F_p\to \{0,\pm 1\}$ is a
quadratic multiplicative character of $\F_p$ defined:
$\smleg{x}{p}$ is $+1$ if $x$ is a square modulo $p$, $-1$ if it
is not a square, and $0$ if $x=0$.  The quantum algorithm of the
previous section showed us how we can determine the shift
$s\in\F_p$ given the function $f_s(x)=\smleg{x+s}{p}$. We now show
how this algorithm enables us to break schemes for `algebraically
homomorphic encryption'.

A cryptosystem is \emph{algebraically homomorphic} if given the
encryption of two plaintexts $E(x)$, $E(y)$ with $x,y\in\F_p$, an
untrusted party can construct the encryption of the plaintexts
$E(x+y)$ and $E(xy)$ in polynomial-time.  More formally, we have
the secret encryption and decryption functions $E:\F_p\to S$ and
$D:S\to \F_p$, in combination with the public add and
multiplication transformations $A:S^2\to S$ and $M:S^2\to S$ such
that $D(A(E(x),E(y)))=x+y$ and $D(M(E(x),E(y)))=xy$ for all
$x,y\in\F_p$.  We assume that the functions $E$, $D$, $A$ and $M$
are deterministic. The decryption function may be many-to-one. As
a result the encryption of a given number can vary depending on
how the number is constructed.  For example, $A(E(4),E(2))$ may
not be equal to $M(E(2),E(3))$.  In addition to the public $A$ and
$M$ functions, we also assume the existence of a zero-tester
$Z:S\to \{0,1\}$, with $Z(E(x))=0$ if $x=0$, and $Z(E(x))=1$
otherwise.

An algebraically homomorphic cryptosystem is a cryptographic
primitive that enables two players to perform noninteractive
secure function evaluation. It is an open problem whether or not
such a cryptosystem can be constructed. We say we can break such a
cryptosystem if, given $E(s)$, we can recover $s$ in time
polylog($p$) with the help of the public functions $A,M$ and $Z$.
The best known classical attack, due to Boneh and
Lipton~\cite{BonehLiptonBBF}, has expected running time
 $O\left(\exp\left(c \sqrt{\log p \log\log p}\right)\right)$
for the field $\F_p$ and is based on a smoothness assumption.

Suppose we are given the ciphertext $E(s)$. Test $E(s)$ using the
$Z$ function. If $s$ is not zero, create the encryption $E(1)$ via
the identity $x^{p-1} \equiv 1 \bmod{p}$, which holds for all
nonzero $x$.  In particular, using $E(s)$ and the $M$ function, we
can use repeated squaring and compute $E(s)^{p-1}=E(1)$ in $\log
p$ steps.

Clearly, from $E(1)$ and the $A$ function we can construct
 $E(x)$ for every $x\in\F_p$.
Then, given such an $E(x)$, we can compute $f(x)=\smleg{x+s}{p}$
in the following way. Add $E(s)$ and $E(x)$, yielding $E(x+s)$,
and then compute the encrypted $(p-1)/2$th
 power\footnote{The Legendre symbol satisfies
 $\smleg{x}{p}=x^{(p-1)/2}$.}
 of $x+s$, giving $E(\smleg{x+s}{p})$.
 Next, add $E(0)$, $E(-1)$ or $E(1)$ and test if it is an
encryption of zero, and return $0$, $1$ or $-1$ accordingly.
Applying this method on a superposition of $\ket{x}$ states,
we can create (after reversibly uncomputing the garbage of
the algorithm) the state
$\smfrac{1}{\sqrt{p-1}}\sum_x f_s(x)\ket{x}$.
We can then recover $s$ by using Algorithm~\ref{alg:finitefield}.
\begin{corollary}
  Given an efficient test to decide if a value is an encryption of
  zero, Algorithm~\ref{alg:finitefield} can be used to break any
  algebraically homomorphic encryption system.
\end{corollary}

We can also break algebraically homomorphic cryptosystems using
Shor's discrete log algorithm as follows. Suppose $g$ is a
generator for $\Fp^*$ and that we are given the unknown ciphertext
$E(g^s)$. Create the superposition
$\sum_{i,j}\ket{i,j,E(g^{si+j})}$ and then append the state
$\ket{\psi_{si+j}}=\sum_t \smleg{g^{si+j}+t}{p} \ket{t}$ to the
superposition in $i,j$ by the procedure described above.  Next,
uncompute the value $E(g^{si+j})$, which gives
$\sum_{i,j}\ket{i,j}\ket{\psi_{si+j}}$.  Rewriting this as
$\sum_{i,r}\ket{i,r-si}\ket{\psi_{r}}$ and observing that the
$\psi_r$ are almost orthogonal, we see that we can apply the
methods used in Shor's discrete log algorithm to recover $s$ and
thus $g^s$.

\section{Shifted Multiplicative Characters of Finite Rings}
\label{sect:rings}

In this section we show how to solve the shifted multiplicative
character problem for $\Zr{n}$ for any completely nontrivial
multiplicative character of the ring $\Zr{n}$ and extend this to
the case when $n$ is unknown. Unlike in the case for finite
fields, the characters may be periodic.  Thus the shift may not be
unique. The Fourier transform is now the familiar Fourier
transform over the additive group $\Zr{n}$.

\subsection{Shifted Multiplicative Characters of $\mathbf{\Zr{n}}$ for Known $\mathbf{n}$}

\begin{definition}\textsc{(Shifted Multiplicative Character Problem
over $\Zr{n}$)} Given $\chi$, a completely nontrivial
multiplicative character of $\Zr{n}$, and a function $f$ for which
there is an $s$ such that $f(x)=\chi(x+s)$ for all $x$. Find all
$t$ satisfying $f(x)=\chi(x+t)$ for all $x$.
\end{definition}
Multiplicative characters of $\Zr{n}$ may be periodic, so to solve
the shifted multiplicative character problem we first find the
period and then we find the shift. If the period is $\ell$ then the
possible shifts will be $\{s, s+\ell, s+2\ell,\dots\}$.

\begin{algorithm}\textsc{(Shifted Multiplicative Character Problem
  over $\Zr{n}$)}
  \ 
  \label{alg:ring}
  \begin{enumerate}
  \item Find the period $\ell$ of $\chi$.  Let $\chi'$ be $\chi$
    restricted to $\{0,\dots, \ell-1\}$.
    \label{alg:ring:period}
    \begin{enumerate}
    \item Create $\sum_{x=0}^{n-1} \chi(x+s) \ket{x}$.
    \item Compute the Fourier transform over $\Zr{n}$ to obtain
      $\sum_{y=0}^{\ell-1} \om{\ell}{-sy} \hat{\chi}'(y) \ket{y n/\ell}$.
    \item Measure \ket{y n/\ell}. Compute $n/\ell=\gcd(n,y n/\ell)$.
    \end{enumerate}
  \item Find $s$ using the period $\ell$ and $\chi'$:
    \label{alg:ring:shift}
    \begin{enumerate}
    \item Create $\sum_{x=0}^{\ell-1} \chi'(x+s) \ket{x}$.
      \label{alg:ring:shift:superposition}
    \item Compute the Fourier transform over $\Zr{\ell }$ to obtain
      $\sum_y \om{\ell }{-sy} \hat{\chi}'(y) \ket{y}$.
    \item For all $y$ coprime to $\ell$,
      $\hat{\chi}'(y)^{-1}$ into the phase to obtain
      $\sum_{y: \hat{\chi}'(y)\neq 0} \om{\ell }{-sy} \ket{y}$.
    \item Compute the inverse Fourier transform
      and measure.
      \label{alg:ring:shift:measure}
    \end{enumerate}
  \end{enumerate}
\end{algorithm}

\begin{theorem}
  Algorithm~\ref{alg:ring} solves the shifted multiplicative
  character problem over $\Zr{n}$ for
  completely nontrivial multiplicative characters of $\Zr{n}$ in
  polynomial time with probability at least $(\smfrac{\phi(n)}{n})^3
  = \Omega((\smfrac{1}{\log\log n})^3)$.
\end{theorem}
\begin{proof}
Note: because $\chi$ is completely nontrivial, $\chi'$ is a
primitive character of $\Z/\ell\Z$.
\begin{enumerate}
 \item
  \begin{enumerate}
   \item
    $\chi(x+s)$ is nonzero exactly when $\gcd(x+s,n)=1$ so by
    Lemma~\ref{lemma:superposition} we can create the superposition
    with probability $\phi(n)/n$.
   \item
    Since $\chi$ has period~$\ell$, the Fourier transform is nonzero
    only on multiples of $n/\ell$.
   \item
    Since $\hat{\chi}'(y) = \overline{\chi'(y)} \hat{\chi}'(1)$,
    and $\chi'(y)$ is nonzero precisely when $\gcd(y,n)=1$, when we
    measure $yn/\ell$ we have $n/\ell=\gcd(n,yn/\ell)$.
   \end{enumerate}
 \item
  \begin{enumerate}
   \item
    Similar to the argument above, we can create the
    superposition with probability $\phi(\ell)/\ell$.
   \item
    The Fourier transform moves the shift~$s$ into the phase.
   \item
    As in the case for the finite field, this can be done by
    computing the phase of $\chi'(y)$ into the phase of \ket{y}.
   \item
    Let $A=\{y \in \Zr{\ell }: \hat{\chi}'(y)\neq 0\}$.
    $A = (\Zr{\ell})^*$ so $|A|=\phi(\ell)$.
    Then the amplitude of \ket{-s} after the Fourier transform is
    $
      \frac{1}{\sqrt{\phi(\ell)}} \frac{1}{\sqrt{\ell }}
        \left( \sum_{y \in A} \om{\ell }{-ys} \, \om{\ell }{ys} \right)
      = \frac{1}{\sqrt{\phi(\ell)}} \frac{1}{\sqrt{\ell }}
        \left( \sum_{y \in A} 1 \right)
      = \sqrt{\frac{\phi(\ell)}{\ell }}.
    $
    So the probability of measuring \ket{-s} is $\phi(\ell)/\ell$.
 \end{enumerate}
\end{enumerate}
  Thus the algorithm succeeds with probability
  $(\phi(n)/n)  (\phi(\ell)/\ell)^2 \geq (\phi(n)/n)^3$,
which in turn is lower bounded by $\Omega((\smfrac{1}{\log\log
n})^3)$.
\end{proof}

\subsection{Shifted Multiplicative Characters of $\mathbf{\Zr{n}}$ for Unknown $\mathbf{n}$}
\label{sect:unknown}

We now consider the case when $n$ is unknown.
\begin{definition}\textsc{(Shifted Multiplicative Character Problem
  over $\Zr{n}$ with Unknown $n$)}~\\
 Given a completely nontrivial multiplicative
 character $\chi:\Zr{n}\to \C$, for some unknown $n$,
 there is an $s$ such that $f(x)=\chi(x+s)$ for all $x$. Find all $t$
 satisfying $f(x)=\chi(x+t)$ for all $x$.
\end{definition}

\begin{theorem}
  Given a lower bound on the size of the period of $f$, we can
  efficiently
  solve the shifted multiplicative character problem over $\Zr{n}$
  for unknown $n$ on a quantum computer.
\end{theorem}

\begin{proof}
  Let $\ell$ be the period of $f$ and $\chi'$ be $\chi$ restricted to
  $\Zr{\ell }$.  Using the Fourier sampling algorithm described in
  Section~\ref{sect:background:FourierSampling}, we can approximately
  Fourier sample $f$ over $\Zr{\ell }$.  Because $\chi'(y)$ is
  nonzero precisely when $\gcd(y,\ell)=1$, this Fourier sampling
  algorithm returns $y/\ell$ with high probability, where $y$ is
  coprime to $\ell$. Thus we can find $\ell$ with high probability.
  Next, apply Algorithm~\ref{alg:ring} to find $s \bmod \ell$.
\end{proof}

\section{The Hidden Coset Problem}

\label{sect:HCP}

In this section we define the hidden coset problem and give an
algorithm for solving the problem for Abelian groups under certain
conditions. The algorithm consists of two parts, identifying the
hidden subgroup and finding a coset representative. Finding a
coset representative can be interpreted as solving a deconvolution
problem.

The algorithms for hidden shift problems and hidden subgroup problems
can be viewed as exploiting different facets of the power of the
quantum Fourier transform. After computing a Fourier transform, the
subgroup structure is captured in the magnitude whereas the shift
structure is captured in the phase. In the hidden subgroup problem we
measure after computing the Fourier transform and so discard
information about shifts. Our algorithms for hidden shift problems do
additional processing to take advantage of the information encoded in
the phase. Thus the solution to the hidden coset problem requires
fully utilizing the abilities of the Fourier transform.

\begin{definition}\textsc{(Hidden Coset Problem)}
  Given functions $f$ and $g$ defined on a group~$G$ such that for
  some $s\in G$, $f(x)=g(x+s)$ for all $x$ in $G$, find the set of all
  $t$ satisfying $f(x)=g(x+t)$ for all $x$ in $G$.  $f$ is given as
  an oracle, and $g$ is known but not necessarily efficiently computable.
\end{definition}

\begin{lemma}
  The answer to the hidden coset problem is a coset of some subgroup
  $H$ of $G$, and $g$ is constant on cosets of $H$.
\end{lemma}

\begin{proof}
  Let $S=\{t \in G: f(x)=g(x+t) \text{ for all } x \in G\}$ be the set
  of all solutions and let $H$ be the largest subgroup of $G$ such
  that $g$ is constant on cosets of $H$.  Clearly this is well defined
  (note $H$ may be the trivial subgroup as in the Shifted Legendre
  Symbol Problem).  Suppose $t_1, t_2$ are in $S$. Then we have
$g(x+(-t_2+t_1))
    =g((x-t_2) + t_1)
    =f(x-t_2)
    =g((x-t_2)+t_2)
    = g(x)$
  for all $x$ in $G$, so $-t_2+t_1$ is in $H$. This shows $S$ is
  a contained in a coset of $H$.  Since $s$ is in $S$ we must have
  that $S$
  is contained in $s+H$. Conversely, suppose $s+h$ is in $s+H$
  (where $h$ is in $H$). Then
  $g(x+s+h)
    =g(x+s)
    =f(x)$
  for all $x$ in $G$, hence $s+h$ is in $S$. It follows that $S=s+H$.
  While this proof was written with additive notation, it carries
  through if the group is nonabelian.
\end{proof}

\subsection{Identifying the Hidden Subgroup}

We start by finding the subgroup $H$. We give two different
algorithms for determining $H$, the ``standard'' algorithm for the
hidden subgroup problem, and the algorithm we used in
Section~\ref{sect:rings}.

In the standard algorithm for the hidden subgroup problem we form
a superposition over all inputs, compute $g(x)$ into a register,
measure the function value, compute the Fourier transform and then
sample. The standard algorithm may fail when $g$ is not distinct
on different cosets of $H$. In such cases, we need other
restrictions on $g$ to be able to find the hidden subgroup $H$
using the standard algorithm. Boneh and
Lipton~\cite{BonehLiptonHiddenLinearFunctions}, Mosca and
Ekert~\cite{MoscaEkert}, and Hales and
Hallgren~\cite{HalesHallgren} have all given criteria under which
the standard hidden subgroup algorithm outputs $H$ even when $g$
is not distinct on different cosets of $H$.

In Section~\ref{sect:rings} we used a different algorithm to
determine $H$ because the function we were considering did not
satisfy the conditions mentioned above. In this algorithm we
compute the value of $g$ into the \emph{amplitude}, Fourier
transform and then sample, whereas in the standard hidden subgroup
algorithm we compute the value of $g$ into a \emph{register}. In
general, this algorithm works when the fraction of values for
which $\hat{g}$ is zero is sufficiently small and the nonzero
values of $\hat{g}$ have constant magnitude.

\subsection{Finding a Coset Representative as a Deconvolution Problem}

Once we have identified $H$, we can find a coset representative by
solving the associated hidden coset problem for $f'$ and $g'$
where $f'$ and $g'$ are defined on the quotient group $G/H$ and
are consistent in the natural way with $f$ and $g$. For notational
convenience we assume that $f$ and $g$ are defined on $G$ and that
$H$ is trivial, that is, the shift is uniquely defined.

The hidden shift problem may be interpreted as a
\emph{deconvolution} problem. In a deconvolution problem, we are
given functions $g$ and $f=g\star h$ (the convolution of $g$ with
some unknown function $h$) and asked to find this $h$. Let
$\delta_y(x)=\delta(x-y)$ be the delta function centered at $y$.
In the hidden shift problem, $f$ is the convolution of
$\delta_{-s}$ and $g$, that is, $f = g\star \delta_{-s}$. Finding
$s$, or equivalently finding $\delta_{-s}$, given $f$ and $g$, is
therefore a deconvolution problem.

Recall that under the Fourier transform convolution becomes
pointwise multiplication. Thus, taking Fourier transforms, we have
$\hat{f} =  \hat{g} \cdot \hat{\delta}_{-s}$ and hence
$\hat{\delta}_{-s}= \hat{g}^{-1} \cdot \hat{f}$ provided $\hat{g}$
is everywhere nonzero. For the multiplication by $\hat{g}^{-1}$ to
be performed efficiently on a quantum computer would require
$\hat{g}$ to have constant magnitude and be everywhere nonzero.
However, even if only a fraction of the values of $\hat{g}$ are
zero we can still approximate division of $\hat{g}$ by only
dividing when $\hat{g}$ is nonzero and doing nothing otherwise.
The zeros of $\hat{g}$ correspond to loss of information about
$\delta_{-s}$.

\begin{algorithm}
\label{alg:HCP:algorithm} \ 
\begin{enumerate}
  \item \label{alg:HCP:superposition}
    Create $\sum_{x \in G} g(x+s)\ket{x}$.

  \item \label{alg:HCP:1stFT}
    Compute the Fourier transform to obtain
    $\sum_{y \in G} \overline{\psi_y(s)}
    \hat{g}(\psi_y) \ket{y}$,
    where $\psi_y$ are the characters of the group $G$.

  \item \label{alg:HCP:invertg}
    For all $\psi_y$ for which $\hat{g}(\psi_y)$ is nonzero
    compute $\hat{g}(\psi_y)^{-1}$ into the phase to obtain
    $\sum_{y,\hat{g}(\psi_y)\neq 0} \overline{\psi_y(s)} \ket{y}$.

  \item \label{alg:HCP:2ndFT} Compute the inverse Fourier transform and
    measure to obtain $-s$.
\end{enumerate}
\end{algorithm}

\begin{theorem}
  \label{algorithm}
  Suppose $f$ and $\hat{g}$ are efficiently computable,
  the magnitude of $f(x)$ is constant for all values of $x$ in $G$ for
  which $f(x)$ is nonzero, and the magnitude of $\hat{g}(\psi_y)$ is
  constant for all values of $\psi_y$ in $\hat{G}$ for which
  $\hat{g}(\psi_y)$ is nonzero.  Let $\alpha$ be the fraction of $x$
  in $G$ for which $f(x)$ is nonzero and $\beta$ be the fraction of
  $\psi_y$ in $\hat{G}$ for which $\hat{g}(\psi_y)$ is nonzero. Then
  the above algorithm outputs $-s$ with probability $\alpha \beta$.
\end{theorem}

\begin{proof}
\begin{enumerate}
 \item
  By Lemma~\ref{lemma:superposition} we can create the superposition
  with probability~$\alpha$.
 \item
  The Fourier transform moves the shift~$s$ into the phase.
 \item
  Because $\hat{g}$ has constant magnitude, for values
  where $\hat{g}$ is nonzero,
  $\hat{g}(\psi_y)^{-1}=C \overline{\hat{g}(\psi_y)}$ for some
  constant $C$. So we can perform this step by computing the phase
  of $\overline{\hat{g}}$ into the phase.
  For the values where $\hat{g}$ is
  zero we can just leave the phase unchanged as those terms are not
  present in the superposition.
 \item
  Let $A=\{y \in G: \hat{g}(\psi_y)\neq 0\}$. Then the amplitude
  of \ket{-s} is
  \begin{align*}
    &\frac{1}{\sqrt{|A|}} \frac{1}{\sqrt{|G|}}
    \left(
    \sum_{y \in A} \overline{\psi_y(s)} \, \overline{\psi_y(-s)}
    \right)
    \\
  &=
    \frac{1}{\sqrt{|A|}} \frac{1}{\sqrt{|G|}}
    \left(
    \sum_{y \in A} 1
    \right)
  =
    \sqrt{\frac{|A|}{|G|}}
  =
    \sqrt{\beta},
  \end{align*}
  so we measure \ket{-s} with probability $\beta$.
\end{enumerate}
  Thus the algorithm succeeds in identifying $s$ with probability
  $\alpha\beta$ and only requires one query of $f$ and one query of
  $\hat{g}$.
\end{proof}

\subsection{Examples}
We show how the hidden shift problems we considered earlier fit
into the framework of the hidden coset problem.  In the shifted
multiplicative character problem over finite fields, $G$ is the
additive group of \Fq, $g=\chi$ and $H$ is trivial since the shift
is unique for nontrivial $\chi$.  In the shifted multiplicative
character problem over $\Zr{n}$, $G$ is the additive group of
$\Zr{n}$, $g=\chi$ and $H$ is the subgroup $\{0,\ell,\dots,n/\ell
\}$, where $\ell$ (which is a factor of $n$) is the period of
$\chi$.  In the shifted period multiplicative character problem
over $\Zr{n}$ for unknown $n$, $G$ is the additive group of $\Z$,
$g=\chi$ and $H$ is the infinite subgroup $\ell\Z$.

\section{Acknowledgments}
We would like to thank the anonymous referee who pointed out the
application of shifted Legendre symbol problem to algebraically
homomorphic cryptosystems and Umesh Vazirani, whose many suggestions
greatly improved this paper.  We also thank Dylan Thurston and an
anonymous referee for pointing out that algebraically homomorphic
cryptosystems can be broken using Shor's algorithm for discrete log.
Thanks to Lisa Hales for helpful last minute suggestions.

\end{document}